\documentclass[aps,prd,onecolumn,groupedaddress,showpacs,nofootinbib,amssymb
]{revtex4}
\usepackage[dvips]{graphicx}
\usepackage{amssymb}
\usepackage{amsmath}
\usepackage{graphicx}
\usepackage{amsfonts}
\usepackage{bm}

\begin{document}

\title{Gauss-Bonnet Gravitational Baryogenesis}
\author{
S.~D.~Odintsov,$^{1,2}$\,\thanks{odintsov@ieec.uab.es}
V.~K.~Oikonomou,$^{3,4}$\,\thanks{v.k.oikonomou1979@gmail.com}}
\affiliation{ $^{1)}$Institut de Ciencies de lEspai (IEEC-CSIC),
Campus UAB, Carrer de Can Magrans, s/n\\
08193 Cerdanyola del Valles, Barcelona, Spain\\
$^{2)}$ ICREA, Passeig LluA­s Companys, 23,
08010 Barcelona, Spain\\
$^{3)}$ Tomsk State Pedagogical University, 634061 Tomsk, Russia\\
$^{4)}$ Laboratory for Theoretical Cosmology, Tomsk State University of Control Systems
and Radioelectronics (TUSUR), 634050 Tomsk, Russia\\
}

\begin{abstract}
In this letter we study some variant forms of gravitational baryogenesis by using higher order terms containing the partial derivative of the Gauss-Bonnet scalar coupled to the baryonic current. This scenario extends the well known theory that uses a similar coupling between the Ricci scalar and the baryonic current. One appealing feature of the scenario we study is that the predicted baryon asymmetry during a radiation domination era is non-zero. We calculate the baryon to entropy ratio for the Gauss-Bonnet term and by using the observational constraints we investigate which are the allowed forms of the $R+F(\mathcal{G})$ gravity controlling the evolution. Also we briefly discuss some alternative higher order terms that can generate a non-zero baryon asymmetry, even in the conformal invariance limit. 
\end{abstract}


\maketitle



\def\pp{{\, \mid \hskip -1.5mm =}}
\def\cL{\mathcal{L}}
\def\be{\begin{equation}}
\def\ee{\end{equation}}
\def\bea{\begin{eqnarray}}
\def\eea{\end{eqnarray}}
\def\tr{\mathrm{tr}\, }
\def\nn{\nonumber \\}
\def\e{\mathrm{e}}

\section{Introduction}

The excess of matter over antimatter in our Universe is one of the unsolved mysteries in cosmology, ever since cosmology became an autonomous research branch. The observational data coming from the Cosmic Microwave Background \cite{bb2}, supported by the Big Bang nucleosynthesis successful predictions \cite{bb1}, indicate an excess of matter over antimatter, and every viable cosmological description should in some way explain this excess in a successful way. One theoretically appealing mechanism for generating the baryon-anti-baryon asymmetry was given in Ref. \cite{baryog1}, which was called as the ``gravitational baryogenesis'' mechanism. Later on, this mechanism was further studied and developed in Refs. \cite{baryog2,baryog3,baryog4,baryog5,baryog6}. The gravitational baryogenesis mechanism makes use of one of the Sakharov  criteria \cite{sakharov}, and the baryon-anti-baryon asymmetry is guaranteed by the presence of a  $\mathcal{C}\mathcal{P}$-violating interaction, which is of the form,
\begin{equation}\label{baryonassterm}
\frac{1}{M_*^2}\int \mathrm{d}^4x\sqrt{-g}(\partial_{\mu} R) J^{\mu}\, .
\end{equation}
The term (\ref{baryonassterm}) can occur in the theory from higher order interactions coming from an underlying effective theory that controls the high energy physics. The parameter $M_*$ in (\ref{baryonassterm}) denotes the cutoff scale of the underlying effective theory, while $J^{\mu}$, $g$ and $R$ stand for the baryonic matter current, the trace of the metric tensor and the Ricci scalar respectively. In effect, for a flat Friedmann-Robertson-Walker (FRW) Universe, the baryon to entropy ratio $\eta_B/s$ is proportional to $\dot{R}$. Notably, in the case that the matter fluid content of the flat FRW is controlled by relativistic matter with equation of state parameter $w=1/3$, the net baryon asymmetry generated by the term (\ref{baryonassterm}) is zero. 

The purpose of this letter is to investigate the consequences of a baryon asymmetry term related to other curvature invariants and specifically related to the Gauss-Bonnet invariant $\mathcal{G}$, which often appears in string-inspired gravities. Also we shall briefly discuss the effect of baryon asymmetry generating terms related to other higher order gravity terms. For the Gauss-Bonnet case, the $\mathcal{C}\mathcal{P}$-violating interaction that will generate the baryon asymmetry of the Universe is of the form, 
\begin{equation}\label{baryonassgausbon}
\frac{1}{M_*^2}\int \mathrm{d}^4x\sqrt{-g}(\partial_{\mu} \mathcal{G}) J^{\mu}\, .
\end{equation}
This kind of terms can possibly occur in higher order gravities coupled with fundamental group fermion currents. As we demonstrate, for the Gauss-Bonnet baryon asymmetry term (\ref{baryonassgausbon}), there are differences in the resulting baryon to entropy ratio and in addition, the latter is non-zero even in the case that the Universe is filled with relativistic matter ($w=1/3$). We shall investigate the cases that the Universe evolution is controlled by a matter fluid with a constant equation of state parameter $w=p/\rho$, and also we investigate the case that the evolution is controlled by an $R+F(\mathcal{G})$ theory of gravity in the presence of a matter fluid with constant equation of state $w$, with $R$ being the Ricci scalar. Finally we shall discuss in brief how the baryon to entropy ratio becomes in the case of higher order gravity terms coupled with fermion currents. 
  
The outline of this letter is as follows: In section II we briefly review the essential features of gravitational baryogenesis and we also investigate in some detail the implication of a Gauss-Bonnet baryogenesis term, by calculating the corresponding baryon to entropy ratio. We discuss the cases that the Universe's evolution is controlled by a perfect fluid and also the case that the evolution is controlled by a $R+F(\mathcal{G})$ theory plus a perfect matter fluid. At the end of section II we briefly discuss the case of higher order gravity gravitational baryogenesis terms and finally, the conclusions follow in the end of the paper.

\section{Gauss Bonnet Baryogenesis}

As we already discussed, the Cosmic Microwave Background observational data \cite{bb2} and also the Big Bang nucleosynthesis predictions \cite{bb1} indicate an excess of matter over antimatter. In addition to these, there are no matter-antimatter annihilation interactions which produce radiation, so this also supports the excess of matter over antimatter. The predicted baryon to entropy ratio is $\frac{\eta_B}{s}\simeq 9.2\times 10^{-11}$ and according to Sakharov \cite{sakharov}, there are three reasons that a baryon asymmetry can occur, firstly, if baryon number violating particle interactions occur, secondly if $\mathcal{C}$ and $\mathcal{C}\mathcal{P}$ violating particle interactions exist, and thirdly if thermodynamical processes in the Universe are non-equilibrium thermodynamical processes. The baryon number violating interactions can in principle be quite slow for generating the observed baryon asymmetry, for example in $SU(5)$ grand unified theories, the proton decay interaction lasts $10^{22}$ billion years, which is almost twice the age of our Universe. Also the non-equilibrium processes are in principle quite difficult to model, so it is more easy to seek for $\mathcal{C}$ and $\mathcal{C}\mathcal{P}$ violating particle interactions. 

 Effectively, in the process of the Universe's expansion, after the temperature of the Universe drops below the critical temperature $T_D$, the remaining asymmetry is approximately equal to \cite{baryog1},
\begin{equation}\label{baryontoentropy}
\frac{n_B}{s}\simeq -\frac{15g_b}{4\pi^2g_*}\frac{\dot{R}}{M_*^2 T}\Big{|}_{T_D}\, ,
\end{equation}
where $g_b$ is the number of the intrinsic degrees of freedom of the baryons. The critical temperature $T_D$ is the temperature of the Universe at which the baryon asymmetry generating interactions occur. 

In the analysis to follow we shall assume that a thermal equilibrium exists, so in all cases which we study, we will assume that the Universe evolves slowly from an equilibrium state to an equilibrium state, with the energy density being related to the temperature $T$ of each state as,
\begin{equation}\label{equilibrium}
\rho=\frac{\pi^2}{30}g_*\, T^4\, ,
\end{equation}
where $g_*$ denotes the number of the degrees of freedom of the effectively massless particles. So for the $\mathcal{C}\mathcal{P}$ violating interaction of Eq. (\ref{baryonassgausbon}), the corresponding baryon to entropy ratio reads,
\begin{equation}\label{baryontoentropygaussbonet}
\frac{n_B}{s}\simeq -\frac{15g_b}{4\pi^2g_*}\frac{\dot{\mathcal{G}}}{M_*^2 T}\Big{|}_{T_D}\, .
\end{equation}
Now depending on the matter content and the theory that controls the evolution, certain differences may occur, which we discuss in the following two sections.

\subsection{The Perfect Fluid Case}

In the standard Einstein-Hilbert gravity framework used in Ref. \cite{baryog1}, if the Universe is filled with a perfect matter fluid with constant equation of state parameter $w=p/\rho$, the Ricci scalar reads,
\begin{equation}\label{ricci}
R=-8\pi G (1-3w)\rho\, ,
\end{equation}
where the Einstein equations are taken into account. Therefore, from Eq. (\ref{ricci}) it can be seen that in the case of a radiation dominated era, the resulting baryon to entropy ratio is zero. In contrast to this, in our case the resulting baryon to entropy ratio is non-zero even in the radiation domination era. Consider a flat FRW background of the form, 
\begin{equation}\label{metricfrw} 
ds^2 = - dt^2 + a(t)^2 \sum_{i=1,2,3}
\left(dx^i\right)^2\, ,
\end{equation}
with $a(t)$ being the scale factor. Assume that the cosmic evolution is described by $a(t)=B\,t^{\beta}$, with $\beta=2/(3(1+w))$, which is generated by an energy density $\rho\sim a^{-3(1+w)}$. In this case, for the FRW metric (\ref{metricfrw}), the Gauss-Bonnet baryon to entropy ratio (\ref{baryontoentropygaussbonet}) reads,
\begin{equation}\label{gbeb1}
\frac{n_B}{s}\simeq \frac{45g_b\,96\,(\beta-1)\beta^{3}}{12\pi^2g_{*}T_D\,M_*^2\,t_D^5}\, ,
\end{equation}
where $t_D$ is the decoupling time corresponding to the critical temperature $T_D$, and also we use the fact that for the flat FRW metric, the Gauss-Bonnet scalar is equal to,
\begin{equation}\label{gausbonehub}
\mathcal{G}=24H^2\left (\dot{H}+H^2 \right ).
\end{equation}
By using the equilibrium equation (\ref{equilibrium}) for the energy density, and also that $\rho=\rho_0\, a^{-3(1+w)}$, the decoupling time as a function of the critical temperature $T_D$ reads,
\begin{equation}\label{tdcritical}
t_D=\left( \frac{\pi^2g_*}{30\rho_0\,B^{4\beta}}\right)T_D^{1/\beta}\, ,
\end{equation}
and therefore the resulting baryon to entropy ratio reads,
\begin{equation}\label{matterfluidbaryontoentropy}
\frac{n_B}{s}\simeq \frac{45g_b\,96\,(\beta-1)\beta^{3}}{12\pi^2g_{*}T_D\,M_*^2}\left( \frac{\pi^2g_*}{30\rho_0\,B^{4\beta}}\right)^{5/4\beta}T_D^{\frac{5}{\beta}-1}\, .
\end{equation}
The radiation domination case corresponds to $\beta=1/2$, and as it can be seen from Eq. (\ref{matterfluidbaryontoentropy}), the resulting baryon to entropy ratio is non-zero, in contrast to the baryon to entropy ratio generated by the term (\ref{baryonassterm}). Depending on the matter content, the ratio (\ref{matterfluidbaryontoentropy}) can be adjusted to satisfy the observational constraints, but the most interesting feature of a perfect fluid dominated Universe for the case of Gauss-Bonnet baryogenesis, is the fact that the ratio is non-zero in the radiation dominated case.   

\subsection{The $R+F(\mathcal{G})$ Case}

\begin{figure}[h] \centering
\includegraphics[width=17pc]{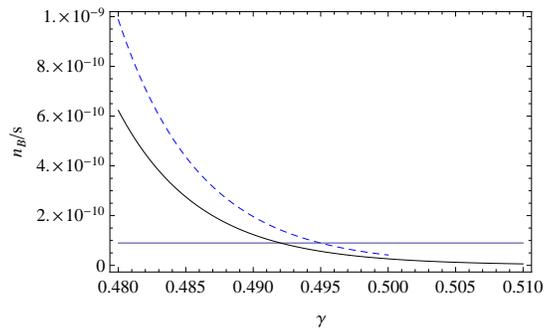}
\caption{The behavior of the baryon to entropy ratio $n_B/s$ as a function of the parameter $\gamma$, for $f_0=10^{15}$, $T_D=2 \times 10^{16}$GeV, $w=1/3$. The blue and dashed curve corresponds to $M_*=10^{12}$GeV, while the black corresponds to $M_*=10^{12.1}$GeV.}
\label{plot1}
\end{figure}

Now we shall calculate the baryon to entropy ratio in the case that the cosmological evolution is governed by an $R+F(\mathcal{G})$ theory of gravity, in the presence of a matter fluid with energy density $\rho_m$ and pressure $p_m$. The gravitational action of the modified gravity theory is in this case \cite{Nojiri:2005jg,Cognola:2006eg},
\begin{equation}\label{actionfggeneral}
\mathcal{S}=\frac{1}{2\kappa^2}\int \mathrm{d}x^4\sqrt{-g}\left [ R+F(\mathcal{G})\right ]+S_m,
\end{equation}
where $\kappa^2=8\pi G$ denotes the gravitational constant and $S_m$ denotes the action of the matter fluids, which in our case consist of a simple perfect fluid with constant equations of state parameter $w=p_m/\rho_m$. By varying the action (\ref{actionfggeneral}) with respect to the metric tensor $g_{\mu \nu}$, we obtain the following equations of motion,
\begin{eqnarray}
\label{fgr1}
&& \!\!\!\!\!\!\!\!\!\!
R_{\mu \nu}-\frac{1}{2}g_{\mu \nu}F(\mathcal{G})+\left(2RR_{\mu \nu}-4R_{\mu 
\rho}R_{\nu}^{\rho}+2R_{\mu}
^{\rho \sigma \tau}R_{\nu \rho \sigma \tau}-4g^{\alpha \rho}g^{\beta \sigma}R_{\mu \alpha 
\nu \beta}
R_{\rho \sigma}\right)F'(\mathcal{G})\notag 
\\ 
&& \  +4 \left[\nabla_{\rho}\nabla_{\nu}F'(\mathcal{G})\right ] R_{\mu}^{\rho}
-4g_{\mu \nu} \left [\nabla_{\rho}\nabla_{\sigma }F'(\mathcal{G})\right ]R^{\rho \sigma }+4 \left 
[\nabla_{\rho}\nabla_{\sigma }F'(\mathcal{G})\right ]g^{\alpha \rho}g^{\beta \sigma }R_{\mu \alpha 
\nu \beta }
\notag 
\\ 
&& \
-2 \left [\nabla_{\mu}\nabla_{\nu}F'(\mathcal{G})\right ]R+2g_{\mu \nu}\left [\square F'(\mathcal{G}) 
\right]R
\notag 
\\
&&\
-4 \left[\square F'(\mathcal{G}) \right ]R_{\mu \nu }+4 
\left[\nabla_{\mu}\nabla_{\nu}F'(\mathcal{G})\right]R_{\nu}^{\rho }
=\kappa^2T_{\mu \nu }^m,
\end{eqnarray}
with $T_{\mu \nu}^m$ being the energy momentum tensor of the matter fluids. In our case $T^{\mu}_{\nu}=\mathrm{diag}(\rho_m,p_m,p_m,p_m)$, and for the flat FRW metric of Eq. (\ref{metricfrw}), the equations of motion (\ref{fgr1}) read,
\begin{equation}
 \label{eqnsfggrav}
  6H^2+F(\mathcal{G})-G\mathcal{G}\,F'(\mathcal{G})+24H^3\dot{\mathcal{G}}F''(\mathcal{G})=2\kappa^2\rho_{m}
\end{equation}
\begin{eqnarray}
 &&4\dot{H}+6H^2+F(\mathcal{G})-G\mathcal{G}\,F'(\mathcal{G})+16H\dot{\mathcal{G}}\left ( \dot{H}+H^2\right ) F''(\mathcal{G})
 \notag\\ 
&&\ \ \ \ \  +8H^2\ddot{\mathcal{G}}F''(\mathcal{G})+8H^2\dot{\mathcal{G}}^2F'''(\mathcal{G})=-2\kappa^2p_{m}.
 \label{eqnsfggrav2}
\end{eqnarray}
Also the matter fluid satisfies the continuity equation $\dot{\rho}_m+3H(\rho_m+p_m)=0$, so effectively $\rho_m=\rho_0a^{-3(1+w)}$. Having these at hand, we shall assume that the functional form of the $F(\mathcal{G})$ gravity is of the form $F(\mathcal{G})=f_0\, \mathcal{G}^{\gamma}$, and in the case that $\gamma <1/2$, it was shown in \cite{Cognola:2006eg} that the cosmological evolution generated by this power-law $F(\mathcal{G})$, has the scale factor $a(t)=B\, t^{\beta}$, with $\beta=4\gamma/(3(1+w))$. By using Eq. (\ref{eqnsfggrav}), replacing the scale factor and the functional form of the $F(\mathcal{G})$ we assumed and finally by keeping only leading order terms, we can find the explicit dependence of the energy density $\rho_m$ as a function of the decoupling time $t_D$, which is,
\begin{equation}\label{expldect}
\rho_m=\mathcal{C}\,t_D^{\gamma}\, ,
\end{equation}
 with $\mathcal{C}=2^{2+3\gamma}3^{\gamma}f_0\,M_p^2\beta^{\gamma}(\beta-1)\gamma\, (\gamma-1)$, where $M_p^2=1/\kappa^2$. Then by combining Eqs. (\ref{equilibrium}) and (\ref{expldect}), we obtain the relation between the decoupling time $t_D$ with the decoupling temperature $T_D$, which is,
\begin{equation}\label{tsdtd}
t_D=(\frac{\pi^2}{30}g_*)^{-\frac{1}{4\gamma}}\,\mathcal{C}^{\frac{1}{4\gamma}}T_D^{\frac{1}{\gamma}}\, .
\end{equation}
Finally, the baryon to entropy ratio for the $F(\mathcal{G})$ case can easily be calculated by combining Eqs. (\ref{baryontoentropygaussbonet}), (\ref{gausbonehub}) and also by taking into account that $F(\mathcal{G})=f_0\, \mathcal{G}^{\gamma}$ and $a(t)=B\, t^{\beta}$, so the result is,
\begin{equation}\label{baryontoentropyratioforthefgcase}
\frac{n_B}{s}\simeq \frac{45g_b96(4\gamma/(3(1+w))-1)(4\gamma/(3(1+w)))^3}{12\pi^2g_*M^2_*(\frac{\pi^2}{30}g_*)^{-\frac{5}{4\gamma}}\mathcal{C}^{5/4\gamma}}T_D^{\frac{5}{\gamma}-1}\, .
\end{equation} 
A direct comparison of the baryon to entropy ratio for the $F(\mathcal{G})$ case to the $F(R)$ case, with $F(R)=f_0R^{\gamma}$, shows that in the $F(\mathcal{G})$ case, the baryon to entropy ratio as a function of the decoupling temperature reads $n_B/s\sim T^{\frac{5}{\gamma}-1}_D$, while in the $F(R)$ case it is $n_B/s\sim T^{\frac{3(1+w)}{4\gamma}-1}_D$.  Hence the two scenarios result to different functional dependencies with respect to the decoupling temperature. Obviously, the observational constraints on the baryon asymmetry can constraint the parameter $\gamma$ or the parameter $f_0$, and therefore the form of the $F(\mathcal{G})$ gravity, as we now show. By assuming that the cutoff scale $M_*$ takes the value $M_*=10^{12}$GeV, also that the critical temperature is equal to $T_D=M_I=2\times 10^{16}$GeV \cite{baryog2}, with $M_I$ being the upper bound for tensor mode fluctuations constraints on the inflationary scale, and finally for $\gamma=0.49$ (recall that $\gamma<1/2$), the parameter $f_0$ has to be $f_0>7.44\times 10^{14}$, in order the constraint $n_B/s<9\times 10^{-11}$ is satisfied. In Table \ref{tablei} we present the allowed values for $f_0$, by using $T_D=M_I=2\times 10^{16}$GeV, $\gamma=0.49$, for various values of the cutoff scale.
\begin{table*}
\small
\caption{\label{tablei} The allowed values of the parameter $f_0$, for various values of the cutoff scale $M_*$.}
\begin{tabular}{@{}crrrrrrrrrrr@{}}
\tableline
\tableline
\tableline
Cutoff Scale & $M_*=10^{11}$GeV & $M_*=10^{12.5}$GeV & $M_*=10^{13}$GeV 
\\\tableline
$f_0$ & $f_0>4.69 \times 10^{15}$ & $f_0>2.94\times 10^{14}$ & $f_0>1.17\times 10^{14}$
\\\tableline
\tableline
 \end{tabular}
\end{table*}  
In addition, in Fig. \ref{plot1}, we plotted the functional dependence of the baryon to entropy ratio $n_B/s$ as a function of the parameter $\gamma$, for $f_0= 10^{15}$, $T_D=2 \times 10^{16}$GeV, $w=1/3$. The blue and dashed curve corresponds to $M_*=10^{12}$GeV, while the black curve corresponds to $M_*=10^{12.1}$GeV. The horizontal line is at the value $n_B/s=9 \times 10^{-11}$, so this is the observational bound. In principle, the parameter $M_*$ can take even lower values from the ones we used, and in effect this would constrain the parameter $f_0$ to take even larger values. 

\subsection{Other Higher Order Gravity Terms}

Some alternative couplings to the one we considered in the previous section contain higher order terms of the Ricci and the Riemann tensor. For example some possible higher order terms containing the Ricci and Riemann tensor, that could potentially generate the baryon asymmetry in the Universe are,
\begin{equation}\label{riccinrewds}
\sim \int \mathrm{d}^4x\sqrt{-g}\,\partial_{\mu}(R^{\alpha \beta}R_{\alpha \beta})J^{\mu},\, \, \,\sim \int \mathrm{d}^4x\sqrt{-g}\,\partial_{\mu}(R^{\alpha \beta\gamma \delta}R_{\alpha \beta\gamma \delta})J^{\mu}
\end{equation}
where again $J^{\mu}$ is the baryonic current. Note that these two terms generate a non-zero baryon to entropy ratio even in the exact conformal case ($w=1/3$), since $\partial_t (R^{\alpha \beta\gamma \delta}R_{\alpha \beta\gamma \delta})\neq 0$, and  $\partial_t (R^{\alpha \beta}R_{\alpha \beta})\neq 0$. In order to demonstrate this, let us briefly calculate the the baryon to entropy ratio for the Ricci tensor containing term. We assume that the background is the flat FRW of Eq. (\ref{metricfrw}), and also that the physics is governed by the standard Einstein-Hilbert gravity. Then, for the power-law cosmology $a(t)=B\,t^{\beta}$, the baryon to entropy ratio is equal to,
\begin{equation}\label{baryontoentropyriemann}
\frac{n_B}{s}\sim \frac{45g_b\,48\,\beta^{2}(1-3\beta+3\beta^2)}{12\pi^2g_{*}\,M_*^2}\left( \frac{\pi^2g_*}{30\rho_0\,B^{4\beta}}\right)^{5/4\beta}T_D^{\frac{5}{\beta}-1}\, .
\end{equation}
Therefore, in this case too, a non-zero baryon asymmetry is predicted for the case $\beta=1/2$, which corresponds to the exact radiation dominated cosmic evolution. Finally, let us note that the ratio of the baryon to entropy ratio for the Gauss-Bonnet baryogenesis term, over the baryon to entropy ratio for the Ricci tensor baryogenesis is,
\begin{equation}\label{ratiocomp}
\frac{\left(\frac{n_B}{s}\right)_{GB}}{\left(\frac{n_B}{s}\right)_{R}}=\frac{96\,(\beta-1)\beta^3}{48\beta^2\,(1-3\beta+3\beta^2)}
\end{equation}
where $\left(\frac{n_B}{s}\right)_{GB}$ and  $\left(\frac{n_B}{s}\right)_{R}$ are the baryon to entropy ratio for the Gauss-Bonnet and Ricci tensor baryogenesis respectively.

\section{Conclusions}

In this letter we studied the gravitational baryogenesis mechanism for the generation of baryon asymmetry by using higher order terms containing the Gauss-Bonnet invariant coupled to the baryonic current. As we demonstrated, in contrast to the standard gravitational baryogenesis mechanism where the derivative of the Ricci  scalar is used, in the Gauss-Bonnet baryogenesis scenario the baryon asymmetry can be generated even in the case that the Universe is dominated by radiation. We calculated the baryon to entropy ratio for the Gauss-bonnet baryogenesis term in the cases that the Universe is described by a standard Einstein-Hilbert gravity and also in the case of an $R+F(\mathcal{G})$ gravity, in the presence of a perfect matter fluid with constant equation of state parameter. In the case of the $R+F(\mathcal{G})$ gravity we investigated the specific case that the $F(\mathcal{G})$ gravity is of the form $F(\mathcal{G})=f_0\mathcal{G}^{\gamma}$, and we calculated the baryon to entropy ratio. For the well known observational bounds for the baryon to entropy ratio it was possible to constrain the values of the parameters $\gamma$ and $f_0$. Finally, we briefly discussed alternative couplings coming from higher order gravities, which contain the Ricci and Riemann tensors.

An extension of this work would be to include higher order terms which contain the derivatives of the functions $f(R_{\mu \nu}R^{\mu \nu})$, $f(R_{\mu \nu \alpha \beta}R^{\mu \nu \alpha \beta})$, which frequently appear in higher order gravities \cite{hog1,hog2}, or string-inspired scalar-Gauss-Bonnet gravities. We hope to address these issues in a future work.

\section*{Acknowledgments}

This work is supported by MINECO (Spain), project
 FIS2013-44881 (S.D.O) and by Min. of Education and Science of Russia (S.D.O
and V.K.O).

\end{document}